\documentclass[compress]{article}
\usepackage{spconf,amsmath,graphicx}
\usepackage{booktabs}
\usepackage{CJKutf8}
\usepackage{tipa}
\usepackage{hyperref}
\usepackage{cleveref}
\usepackage[numbers,sort&compress]{natbib}
\usepackage{color}
\usepackage{amsfonts}
\usepackage{multirow}
\usepackage{amssymb}
\usepackage{array}
\usepackage{threeparttable}
\usepackage{etoolbox}

\newcommand{\PreserveBackslash}[1]{\let\temp=\\#1\let\\=\temp}
\newcolumntype{C}[1]{>{\PreserveBackslash\centering}p{#1}}
\newcolumntype{R}[1]{>{\PreserveBackslash\raggedleft}p{#1}}
\newcolumntype{L}[1]{>{\PreserveBackslash\raggedright}p{#1}}

\makeatletter
\newcommand{\namedlabel}[2]{%
  \@bsphack
  \protected@write\@auxout{}{%
    \string\newlabel{#1}{%
        {\@currentlabel}%
        {\thepage}%
        {{#2}}
        {\@currentHref}{}%
      }%
  }%
  \@bsphack
}%
\makeatother

\title{LMEC: Learnable Multiplicative Absolute Position Embedding Based Conformer for Speech Recognition}
\usepackage{authblk}

\author[1]{Yuguang Yang}
\author[2]{Yu Pan}
\author[1]{Jingjing Yin}
\author[1]{Heng Lu}
\affil[1]{Ximalaya Inc., ShangHai, China}
\affil[2]{University of Alberta, Edmonton, Canada}

\begin{document}

\maketitle

\begin{abstract}

This paper proposes a Learnable Multiplicative absolute position Embedding based Conformer (LMEC).
It contains a kernelized linear attention (LA) module called LMLA to solve the time-consuming problem for long sequence speech recognition as well as an alternative to the FFN structure.
First, the ELU function is adopted as the kernel function of our proposed LA module.
Second, we propose a novel Learnable Multiplicative Absolute Position Embedding (LM-APE) based re-weighting mechanism that can reduce the well-known quadratic temporal-space complexity of softmax self-attention.
Third, we use Gated Linear Units (GLU) to substitute the Feed Forward Network (FFN) for better performance.
Extensive experiments have been conducted on the public LibriSpeech datasets.
Compared to the Conformer model with cosFormer style linear attention, our proposed method can achieve up to 0.63\% word-error-rate improvement on test-other and improve the inference speed by up to 13\% (left product) and 33\% (right product) on the LA module.
\end{abstract}


\vspace{-2mm}
\section{Introduction}
Over the last decade, end-to-end models have been applied as mainstream approaches for state-of-the-art automatic speech recognition (ASR) systems.
One major architecture is attention based encoder-decoder (AED)\cite{wu2021u2++,vanilla_attention,chan2016listen,dong2018speech}.
A recent trend of building an encoder-decoder ASR model is using stacks of Transformer block or its variations with joint cross entropy and Connectionist Temporal Classification (CTC) loss\cite{ctc}.

Compared with the network of vanilla transformer structure, the Conformer\cite{gulati2020conformer} block combines the advantages of convolution's ability to capture local information and the transformer's ability to observe global features.
Furthermore, since the transformer structure lacks the capture of time series
information, position embedding is used, like relative position encoding\cite{transformer-xl}(XL-RPE), absolute position embedding (APE)\cite{vanilla_attention}, and so forth.
Generally, XL-RPE performs better than APE in the Conformer, especially on ASR tasks.

However, the Conformer block still has some critical points that can be improved.
The main drawback is the quadratic temporal-space complexity due to the core softmax self-attention operation, which is especially important for long input sequences.
In addition, the Feed Forward Network (FFN) layer can also be optimized, as it occupies most of the computation in the Conformer block and plays a decisive role in the effect of the model performance.

Many excellent works\cite{tay2020efficient} have been proposed to tackle the high computational and memory cost in Transformer-style blocks.
The bottleneck mainly comes from the calculation of $O(N^2)$ in self-attention, where $N$ refers to the length of the input sequence.
To be specific, in order to reduce the amount of calculation while minimizing quality loss, some works focus on sparse attention maps or through low-rank decomposition such as\cite{su2021roformer,wang2020linformer,huang2019interlaced}.
Meanwhile, some other works' direction is to turn the calculation of the original left product into a right product\cite{huang2019ccnet,zhen2021cosformer,choromanski2020performer}.
Besides, \cite{tay2005synthesizer} proposes a synthetic self-attention module to approximate attention weights.
Efficient Conformer\cite{burchi2021efficient} try to use grouped multi-head attention to reduce its complexity.
Recently, kernel based self-attention\cite{zhen2021cosformer,choromanski2020performer,linear_attention1,linear_attention2, peng2021random}
has demonstrated its advantages over vanilla transformers, especially in long sequence modeling and inference speed.

In this paper, inspired by cosFormer\cite{zhen2021cosformer}, Performer\cite{choromanski2020performer} and RFA\cite{peng2021random},
which decouple the $O(N^2d)$ operation to $O(Nd^2)$ by replacing self-attention calculation from $(QK^T)V$ to $Q(K^TV)$, we hope to realize this kind of self-attention decoupling in ASR Conformer block.
In the actual inference phase for the ASR task, the audio input range may vary from a few seconds to tens of seconds, which will cause the model to fluctuate in the amount of calculation and memory usage.
Due to the complexity of $O(N^2)$ operation, it will make the inferential computing of long audio more difficult.

Moreover, such decoupling for self-attention somewhat conflicts with the implementation of XL-RPE in the Conformer block, because the scale product of $Q$ and $K$ as well as relative shift operation cannot be applied to the linear complexity output $Q(K^TV)$.
Several attempts have been made to introduce other RPE to the LA module\cite{su2021roformer,zhen2021cosformer}.
Nevertheless, these complex position embedding approaches are not that cost-effective in ASR tasks.

Another concern in the Conformer block is the Macaron-style FFN module, which usually consists of two fully connected layers and an activation function.
Although its infrastructure is simple, there are still some works to optimize this part through sharing parameters like grouped feed forward module\cite{jiang2020convbert} or changing the computing paradigm like\cite{shazeer2020glu}.

The main contributions of our work are as below:
\begin{itemize}
\item We propose a new linear attention kernel, which is much lighter than conformer self-attention and cosFormer attention. We apply this kernel to the Conformer and test it on the ASR task.
\item We propose to apply the Gated Linear Units layer instead of the Feed Forward Network in Conformer block.
\item We recommend different computational strategies for training and inference.
Left product is used during training to ensure model quality.
Dynamically choosing the product mechanism ensures the inference speed during inference.
\end{itemize}
To sum up, our goal is to design a model architecture with better performance in ASR tasks, lower latency in actual use, and better overall cost performance than Conformer.


\vspace{-2mm}
\section{Related Work}
Since transformer is widely used, many workers in classic tasks hope to have a better self-attention paradigm.
For example, \cite{li2021efficient} and \cite{sun2022locality} both tried to design a new linear attention kernel and applied it on the ASR task.
Their aims on linear attention (LA) part are the same as ours, which is decoupling the self-attention operation from attention score calculation
and changing computational complexity from $O(N^2d)$ to $O(Nd^2)$.

\vspace{-2mm}
\subsection{Transformer Block}
Let's review the structure of the standard vanilla transformer block in Fig.\ref{fig:LMLA} (a), which is a stack of self-attention module with feed forward layer.

The key structure of transformer is undoubtedly self-attention module.
Suppose hidden features from the previous layer is $x$,
it will be projected into $Q=xW_q$, $K=xW_k$ and $V=xW_v$ matrix.
Then, self-attention output can be represented as below:

\begin{equation}\label{sa:vanilla}
    \setlength{\abovedisplayskip}{-2mm}
    O_{attn}=SA(Q,K,V)=\textit{softmax}(\frac{QK^\intercal}{\sqrt{d_k}})V
\end{equation}
where $SA(\cdot)$ stands for self-attention operation, Q, K and V stands for query, key and value matrix, $d_k$ is the hidden dimension.


\subsection{Conformer Block}
Combining the transformer's advantages in capturing global information with CNN's advantages in capturing local information, the Conformer block has made great achievements and has already become a benchmark in speech recognition tasks.

\begin{figure*}[pt]
	\centering
	\includegraphics[height=8cm]{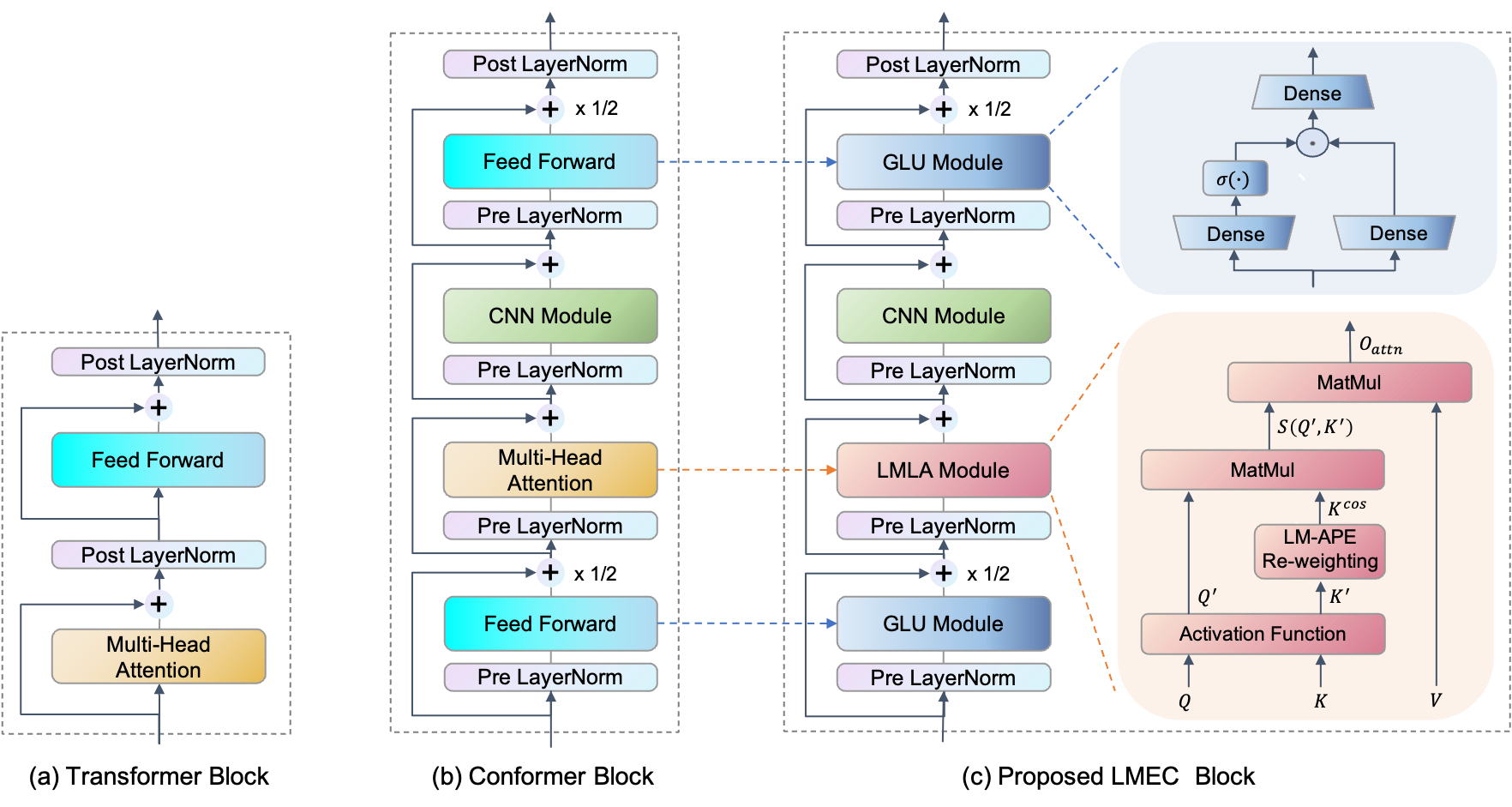}
	\caption{\label{fig:LMLA} The whole structure of Our Mentioned Blocks}
\end{figure*}

As shown in Fig.\ref{fig:LMLA} (b), Conformer block\cite{gulati2020conformer} is a stack of feed forward module, multi-head attention module, convolution module and feed forward module again.
Suppose the input of $i$-th layer is $x_i$, the specific calculation method is as follows:

\begin{equation}\label{eq:conformer}
    \setlength{\abovedisplayskip}{-2mm}
    \begin{split}
         y_i=LN^{\tiny{\textcircled{+}}}( \frac{1}{2} FFN^{\tiny{\textcircled{+}}}( \text{Conv}^{\tiny{\textcircled{+}}}(SA^{\tiny{\textcircled{+}}}( \frac{1}{2} FFN^{\tiny{\textcircled{+}}}(x_i) ) ) )
    \end{split}
\end{equation}
Here the residual function is expressed as $F^{\tiny{\textcircled{+}}}(x) = x + F(x)$.

It is easy to see from the figure that self-attention and feed forward modules are the most computationally intensive net-work structures, which we will focus on later in this paper.

\subsection{cosFormer Kernel}


A self-attention that can disassemble left and right multiplication has been redesigned in cosFormer\cite{zhen2021cosformer}.
In cosFormer, ReLU linear attention activation and cos-based re-weighting mechanism is used as the substitute for softmax operator.
And the row-wise similarity function of $Q$ and $K$ is expressed as:
\begin{equation}\label{eq:cosformer1}
    s(Q_i^\prime,K_j^\prime)=\psi(Q_i)\psi(K_j)cos(\frac{\pi}{2}\times\frac{i-j}{M})
\end{equation}
where $\psi(\cdot)$ stands for linear attention activation operation.
And the formulation can be decomposed as:
\begin{equation}\label{eq:cosformer-decomposed}
    \begin{split}
        s(Q_i^\prime,K_j^\prime)=&(\psi(Q_i)cos(\frac{\pi i}{2M}))(\psi(K_j)cos(\frac{\pi j}{2M}))^{\intercal} \\
        +&(\psi(Q_i)sin(\frac{\pi i}{2M}))(\psi(K_j)sin(\frac{\pi j}{2M}))^{\intercal}
    \end{split}
\end{equation}
Then the new version of self-attention operation could be derivatived from Eq.\ref{sa:vanilla} to
\begin{equation}\label{eq:cosformer2}
\begin{split}
        O_{attn}=&S(Q,K)V \\
        =&(Q^{cos}K^{cos}+Q^{sin}K^{sin})V \\
        =&Q^{cos}(K^{cos}V)+Q^{sin}(K^{sin}V)
    \end{split}
\end{equation}


This approach eliminates the computation bottleneck of long sequence $O(N^2)$ calculation compared to the vanilla transformer through clever formula disassembly and proves its effectiveness in NLP tasks.

From another perspective, this cos-based multiplicative item in cosFormer realizes a new form of RPE.
In order to facilitate the subsequent comparation, we named the cosFormer kernel as a multiplicative relative position embedding (M-RPE) LA.

\vspace{-2mm}

\vspace{-2mm}
\section{Proposed Methodology}

In this section, we propose a Conformer-style block called LMEC based on kernelized LA and GLU module.
First, we focus on designing our linear attention kernel by choosing an appropriate activate function and improving the re-weighting mechanism.
The primary purpose is to acquire the concentration ability for the attention weights' distribution, just retaining the property of softmax attention, but with a much more straightforward calculation.
Then, we introduce variations of Additive RPE and No PE LA to fully illustrate the effect of different positional embedding.
Next, we adopt GLU module and use it to substitute FFN layer of Conformer block.
Finally, we explore the differences between left product and right product training.
The overall architecture of our LMLA module is shown Fig.\ref{fig:LMLA} (c).

\vspace{-4mm}
\subsection{MLA: M-APE Based Linear Attention Kernel}

Here we start with two most important points in MLA, the selection of Linear Attention Activation and the Re-weighting mechanism.
\vspace{-6mm}
\subsubsection{Linear Attention Activation Function}


Following the decompositon approach in cosFormer\cite{zhen2021cosformer}, we adopt the decomposable kernel function to rewrite the attention matirx as:
\begin{equation}\label{eq:npe-la}
    s(Q_i^\prime,K_j^\prime)=\psi(Q_i)\psi(K_j)
\end{equation}
where $\psi(\cdot)$ maps each row of $Q$ and $K$ to their hidden representations $Q_i^\prime$ and $K_j^\prime$.

However, what kind of mapping method can achieve a better effect in ASR task is worth exploring.
The previous experience of scholars in the NLP task is that it is more beneficial to the performance and convergence of the model after mapping the input hidden representation matrix from the upper layer to the non-negative range\cite{zhen2021cosformer,peng2021random,NIPS2016_814a9c18,https://doi.org/10.48550/arxiv.1704.00805,choromanski2020performer}.
We have selected four common activation functions as our kernel function for experimental comparison:
\begin{equation}
    \begin{split}
        \psi_{relu}(x)    &= Relu(x) \\
        \psi_{sigmoid}(x) &= Sigmoid(x) \\
        \psi_{tanh}(x)    &= 0.5 + Tanh(x) + 0.5 \\
        \psi_{elu}(x)     &= ELU(x) + 1 \\
    \end{split}
\end{equation}

\subsubsection{M-APE Based Re-weighting Mechanism} 

Expected to improve the re-weight mechanism, we make some efforts in introducing diverse RPE forms to the decomposed LA kernel. But it seems not cost-effective for ASR tasks.
Results of different positional embedding we will illustrate in later experiments.
Here, we concentrate on how we optimize cosFormer RPE, as the twice $QK^\intercal$ style calculation in which still has redundant parts.

We degenerate original Eq.\ref{eq:cosformer1} from relative position coding into absolute position coding and is expressed as follows:
\begin{equation}\label{eq:abs-la}
    s(Q_i^\prime,K_j^\prime)=\psi(Q_i)\psi(K_j)cos(\frac{\pi}{2}\times\frac{j}{M})
\end{equation}
which full name is called multiplicative absolute positional embedding linear attention (M-APE) LA.

\subsection{LMLA: LM-APE Based Linear Attention Kernel} 
Furthermore, some works\cite{devlin2018bert, T5} show that random learnable position vectors could achieve better results on specific tasks.
Therefore, we simplify the Eq.\ref{eq:abs-la} as:

\begin{equation}\label{eq:lm-ape}
    s(Q_i^\prime,K_j^\prime)=\psi(Q_i)\psi(K_j)cos(R_j)
\end{equation}
where $R$ stands for a random vector.
$R_j$ is a learnable value for each position $j$ only applied to the key matrix, so that there is no need to calculate the scale product operation of QK twice as in Eq.\ref{eq:cosformer2}.
Uniformly, we call this linear attention with learnable multiplicative absolute positional embedding (LM-APE) as LMLA.

\begin{figure}[h]
	\centering
	\includegraphics[width=8cm]{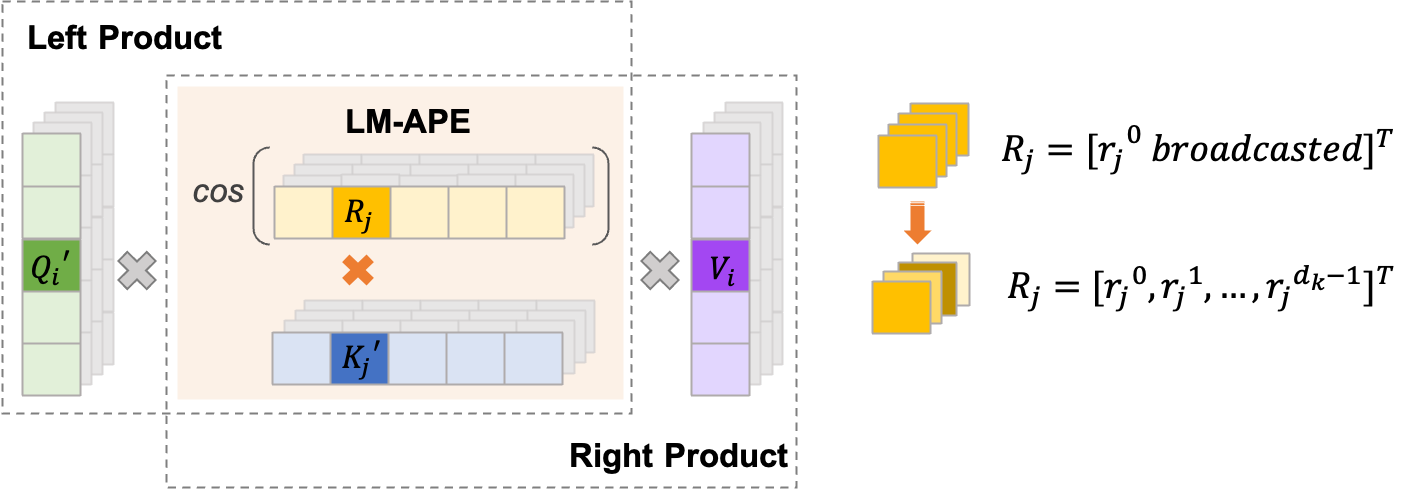}
	\caption{\label{fig:lmla} Explaination of Re-weighting Mechanism in LMLA}
\end{figure}


As mentioned above, our proposed APE-based re-weighting approaches are more straightforward and faster than the M-RPE of cosFormer in Eq.\ref{eq:cosformer-decomposed}.
To improve the model's expressiveness, we extend the embedding of a specific position $j$ from a value to a vector as compensation, slightly increasing the parameters and retaining the calculation cost.
Fig\ref{fig:lmla} illustrates how LM-APE in Eq.\ref{eq:lm-ape} is applied to each element $K_j^\prime \in \mathbb R^{1\times d_k}$.
If $R_j \in \mathbb R^{1\times 1}$, it will be automatically broadcasted to $\mathbb R^{1 \times d_k}$ before element-wise production of $\psi(K_j)cos(R_j)$.
In our design, $R_j=[r^0, r^1, \dots, r^{d_k-1}]^\intercal$, where each $r$ is individually learnable.
For M-APE, we use a simple affine transform $W_{ext} \in \mathbb{R}^{1 \times d_k}$ to extend $\cos(\frac{\pi}{2}\times\frac{j}{M})$ mentioned in Eq.\ref{eq:abs-la} to a $d_k$ dimension vector for making the comparison fair. Our subsequent experimental conclusions are also adopted in this way by default.

\vspace{-2.5mm}
\subsection{Variations for Comparison}

\subsubsection{Additive Relative Position Embedding Linear Attention} 

In order to involve positional information by product operation, we also tried to add information through bias.
Besides, the basic calculation of self-attention with relative position embedding (XL-RPE), which is slightly different from Eq.\ref{sa:vanilla} is as below:
\begin{equation}\label{sa: vanilla}
  SA(Q,K,V)=\textit{softmax}(\frac{QK^\intercal+S_{bd}}{\sqrt{d_k}})V
\end{equation}
where $S_{bd}$ stands for the $b$ and $d$ option from transformer-xl, which is actually calculated by $Q$ and linear mapped position encoding with relative shift operation.
Due to space limitations, those interested in specific details can refer to \cite{transformer-xl}.
Here, we derive the value of $S_{bd}$ to $\cos(\frac{\pi} {2} \times \frac {i-j} {M})$ in order to decouple $QK$ calculation.
Here, we call it additive relative positional embedding (A-RPE) LA module.

\subsubsection{No Position Embedding Linear Attention} 

We hypothesize there is no performance degradation after changing the position embedding method from RPE into APE in the linear attention paradigm,
In order to prove the effectiveness of APE, we also compare the variation of linear attention with no positional embedding (NPE).
On the basis of our proposed formula, if the positional embedding part is removed, its formula can be expressed as Eq.\ref{eq:npe-la},
where other settings are totally the same as LMLA block except for the position embedding.

\subsection{Gated Linear Units}

The Conformer block usually applies Convolutional Neuron Network (CNN) layer followed by Feed Forward Network (FFN) layer,
which is represented as below (here we omit bias representation):
\begin{equation}
    O_{ffn}=\sigma(xW_1)W_2
\end{equation}
where $\sigma(\cdot)$ stands for activation functions.

Besides, $W_1 \in \mathbb{R}^{h_o \times h_{ffn}}$ and $W_2 \in \mathbb{R}^{h_{ffn} \times h_o}$,
where $h_o$ is the hidden dimension of encoder output, $h_{ffn}$ is the hidden dimension of FFN layer.
We propose to replace the FFN layer with the Gated Linear Units layer\cite{shazeer2020glu}, which represents as below:
\begin{equation}
    O_{glu}=(\sigma(xW_1)\otimes xW_2)W_3
\end{equation}
where $\otimes$ stands for component-wise product operation,
$W_1, W_2 \in \mathbb{R}^{h_o \times h_{glu}}$, $W_3 \in \mathbb{R}^{h_{glu} \times h_o}$.
For keeping the number of parameters unchanged, we set $h_{glu}=\frac{2} {3} h_{ffn}$.

\subsection{Training and Inference Method}
In linear attention designed by us, the results of left product (L-Prod) calculations and right product (R-Prod) calculations are consistent.
But in the actual training process, the gradients update differently.
Suppose the simplified LMLA formula of left product and right product training is as below:
\begin{equation}
        O_{attn} = \underbrace{(\textcolor{red}{Q^\prime} (\textcolor{red}{K^\prime \otimes R^{cos}})) V}_{\textcolor{red}{L-Prod}} = \underbrace{Q^\prime ((\textcolor{blue}{K^\prime \otimes R^{cos}}) \textcolor{blue}{V})}_{\textcolor{blue}{R-Prod}}
\end{equation}
%
where $Q^\prime=\psi (Q)$, $K^\prime=\psi (K)$, $R^{cos}=cos(R)$.

Based on our experiments, we suggest an L-Prod approach for training and a dynamic chosen inference method.
In other words, in the training stage, we try to use L-Prod calculation to ensure training effect and convergence stability.
When $N \le d$, we use L-Prod inference, and when $N \ge d$, we use R-Prod inference to maximize the inference speed.

\vspace{-2mm}
\section{Experiments}
\begin{table*}[htbp]\small
    \vspace{-4mm}
    \centering
    \caption{WER(\%) results on LibriSpeech for different models}
    \begin{tabular}{cccC{6mm}C{12mm}C{6mm}C{20mm}C{6mm}C{20mm}} 
        \hline
        \toprule \\
        \specialrule{0em}{-10pt}{1pt}
        \multirow{2}{*}[-3mm]{Model} & \multirow{2}{*}[-3mm]{Activation} & \multirow{2}{*}[-3mm]{LA Style} & \multirow{2}{6mm}[-3mm]{with GLU} & \multirow{2}{12mm}[-3mm]{Heads} & \multicolumn{2}{c}{Test Clean} & \multicolumn{2}{c}{Test Other} \\
        \cmidrule{6-9}  & &  &  &  &  ctc greedy & attention rescore & ctc greedy & attention rescore   \\
        \midrule
        Conformer0\cite{gulati2020conformer} & -      & XL-RPE  & $\times$     & 4  & \textcolor{blue}{3.64}  & \textcolor{blue}{3.26}  & \textcolor{blue}{9.28}  & \textcolor{blue}{8.51} \\
        Conformer1\cite{gulati2020conformer} & -      & XL-RPE  & $\times$     & 8  & \textcolor{blue}{3.51}  & \textcolor{blue}{3.31}  & \textcolor{blue}{9.12}  & \textcolor{blue}{8.51} \\
        \midrule
        cosFormer0\cite{zhen2021cosformer}   & ReLU     & M-RPE-LA  & $\times$   & 4  & 3.79  & 3.38  & 9.80  & 8.95 \\
        cosFormer1\cite{zhen2021cosformer}   & ReLU     & M-RPE-LA  & $\times$   & 8  & 3.85  & 3.44  & 9.80  & 9.00  \\
        \midrule
        LBLA0\cite{sun2022locality}         & Sigmoid  & M-RPE-LA  & $\times$   & 4  & 3.63     & 3.19     & 9.35     & 8.51      \\
        LBLA1\cite{sun2022locality}         & Sigmoid  & M-RPE-LA  & $\times$   & 8  & 3.60  & 3.23  & 9.38  & 8.68   \\
        \midrule
        LMEC0 (Ours)       & ELU      & LM-APE-LA &\checkmark   & 4  & \textbf{3.54}  & \textbf{3.20}  & \textbf{9.13}  & \textbf{8.32}   \\
        LMEC1 (Ours)       & ELU      & LM-APE-LA &\checkmark   & 8  & 3.57  & \textbf{3.17}  & 9.18  & \textbf{8.38}  \\
        \bottomrule
    \end{tabular}\label{tb:all}
    \begin{tablenotes}\footnotesize
        \item \textbf{Model\{0/1\}}: Model with 4 or 8 attention head on attention module;
        \textbf{\{cosFormer/LBLA\}\{0/1\}}: Conformer with cosFormer/LBLA attention module; \\
    \end{tablenotes}
    \vspace{-4mm}
\end{table*}

\subsection{Experimental Dataset \& Settings}
In this work, we evaluate our proposed model and other state-of-the-art models on the open source dataset LibriSpeech\cite{2015Librispeech}.
LibriSpeech consists of 1000 hours of 16kHZ labeled English audio, which is divided into train, dev and test.
We train all models on the LibriSpeech training dataset which contains approximately 960 hours of 16kHZ English speech with its corresponding text, and evaluate them on the LibriSpeech test-clean and test-other datasets.


We first use SentencePiece\cite{2018SentencePiece} to build a byte pair encoding tokenizer, which generates 5000 subword pieces from the transcripts of LibriSpeech.
And the audios are converted to 80-dimensional filter-bank feature sequences, and Spec-Augment\cite{2019SpecAugment} is used during training.
All models are trained with 120 epochs, and optimized by Adamw\cite{adamw}, where the original learning rate is 0.001 and warm up step is 10000 with Cosine Annealing\cite{Cosine_Annealing} scheduler.
In terms of model structure for encoder, 12 layers of the Conformer style block is used, in which convolutional kernel size is 15, model dimension is 256, hidden parameter of FFN layer is 2048.
The decoder has 6 layers single directional transformer with 4 attention heads and FFN dimension is 2048.
All our experiments are conducted on WeNet\cite{wu2021u2++} Toolkit and uniformly trained under the condition of 4 A100 GPUs.

\vspace{-4mm}
\subsection{Main Results}
\vspace{-2mm}
%

In this section, we compare the overall performance of our proposed LMEC model with some state-of-the-art Conformer-style models, including baseline Conformer\cite{gulati2020conformer}, cosFormer\cite{zhen2021cosformer}, LBLA\cite{sun2022locality}.
For sufficient comparison, we compare 4 results between our model and the references on LibriSpeech test-clean/test-other.
Two most common attention heads (4 / 8) and two evaluation methods (ctc greedy search / attention rescoring) are adopted in our experiments.

Compared with Conformer, our proposed LMEC has 3 out of 4 results that perform better, achieving 0.2$\sim$4.2\% relative WER reduction on test-clean.
Only one performs slightly worse.
And similar conclusions can be achieved on test-other.
More, the linearized attention kernel (LMLA) can be computed more easily and quickly than the softmax attention.

In contrast to cosFormer and LBLA, which are also linear kernels, our performance gains are more significant.
Compared with cosFormer, LMLA can achieve up to 5.3$\sim$7.8\% relative improvement on test-clean and 6.3$\sim$7.0\% on test-other.
LMLA also shows 0.1$\sim$2.5\% (test-clean) and 2.13$\sim$3.46\% (test-other) relative promotion than LBLA.
All the results validate the effectiveness of the proposed LMEC model.

\begin{figure}[h]
    \vspace{-2mm}
    \setlength{\abovedisplayskip}{-5mm}
	\centering
	\includegraphics[height=4.2cm]{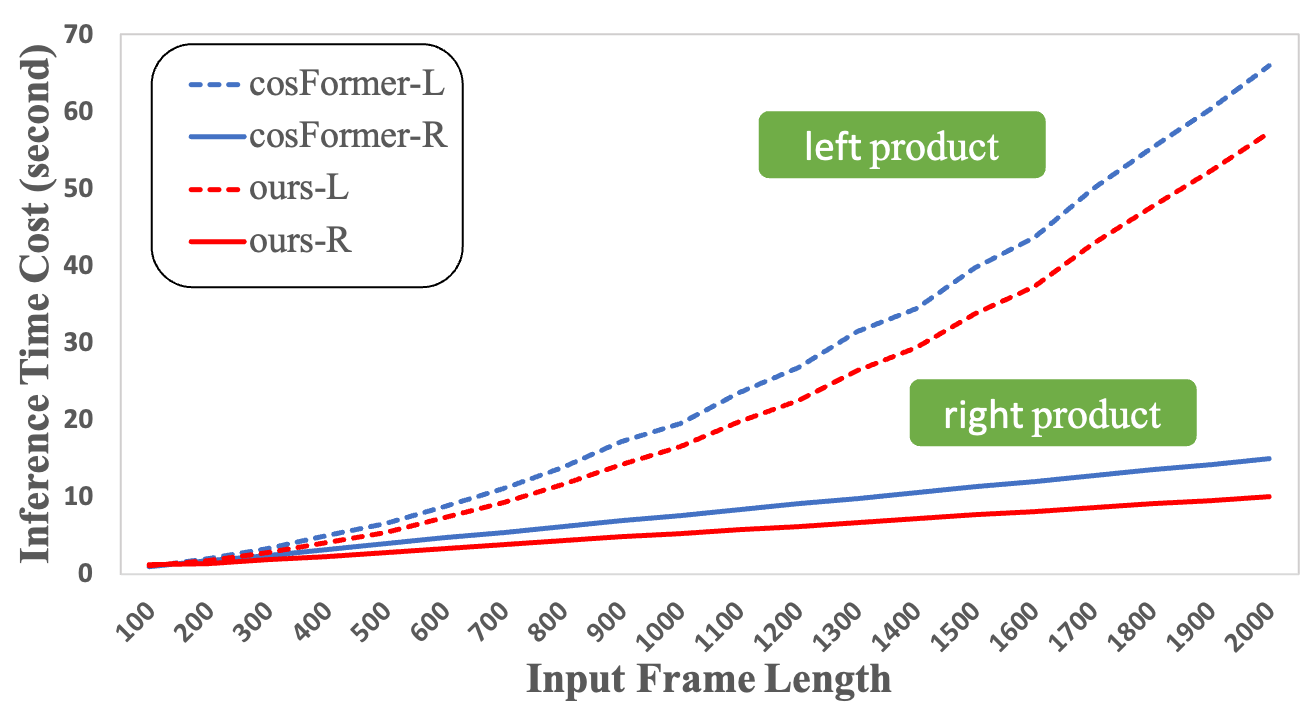}
	\caption{\label{fig:cmp2} Inference Time Cost of cosFormer and our LMLA}
    \setlength{\belowdisplayskip}{-5mm}
\end{figure}

We use the input signal from 100 to 2000 frames (each frame stands for 40ms audio) with the batch size of 100 on a single A100 GPU to test the inference.
All tests adopted GPU warm-up inference 1000 times and averaged.
One of the most significant features of Fig.\ref{fig:cmp2} is that as the input sequence gets longer, the speed advantage of right-product computing becomes more evident than left-product computing.
Compared with cosFormer, LMLA reduces the averaged left-product time cost from 66s to 57s (rel -13\%) and the right-product time cost from 15s to 10s (rel -33\%) when the input sequence length is 2000 (80s).

In conclusion, our proposed model LMLA obtains better performance and fewer inference costs than Conformer, cosFormer and LBLA.

\vspace{-2mm}
\subsection{Ablation Studies}
\subsubsection{Effect of Our Proposed LA Module}
In this section, we evaluate the effectiveness of our proposed
\begin{table}[htbp]\small
    \vspace{-4.5mm}
    \centering
    \caption{\label{tb:la-compare} Effect of Our Proposed LA Module}
    \begin{tabular}{C{13mm}C{8mm}C{12mm}C{5mm}C{8.5mm}C{5mm}C{8.5mm}}
        \hline
        \toprule\\
        \specialrule{0em}{-10pt}{1pt}
        \multirow{2}{13mm}[-3mm]{Model} & \multirow{2}{8mm}[-1mm]{LA Acti-vation} & \multirow{2}{12mm}[-3mm]{LA Style} & \multicolumn{2}{c}{Test Clean} & \multicolumn{2}{c}{Test Other} \\
        \cmidrule{4-7}  &  &  &  ctc greedy & attention rescore & ctc greedy & attention rescore   \\
        \midrule
        \textbf{LMLA-R}       & ReLU      & LM-APE   & 3.74  & 3.29  & 9.37  & 9.05 \\ 
        \textbf{LMLA-T}       & Tanh      & LM-APE   & 3.78  & 3.33  & 9.73  & 8.91 \\
        \textbf{LMLA-S}       & Sigmoid   & LM-APE   & \textbf{3.67}  & \textbf{3.28}  & 9.37  & 8.51 \\
        \textbf{LMLA-E}       & ELU       & LM-APE   & 3.70  & 3.32  & \textbf{9.37}  & \textbf{8.47} \\
        \midrule
        MLA-R        & ReLU      & M-APE    & 3.97  & 3.53  & 9.87  & 9.04 \\ 
        MLA-T        & Tanh      & M-APE    & 3.91  & 3.48  & 9.63  & 8.79 \\
        MLA-S        & Sigmoid   & M-APE    & \textbf{3.73}  & \textbf{3.29}  & 9.43  & 8.60 \\
        MLA-E        & ELU       & M-APE    & 3.80  & 3.39  & \textbf{9.36}  & \textbf{8.52} \\
        \midrule
        V0-R        & ReLU      & A-RPE    & 3.79  & 3.35  & \textbf{9.62}  & 8.82 \\ 
        V0-T        & Tanh      & A-RPE    & 3.85  & 3.39  & 9.73  & 8.82 \\
        V0-S        & Sigmoid   & A-RPE    & 3.90  & 3.44  & 9.97  & 9.06 \\
        V0-E        & ELU       & A-RPE    & \textbf{3.68}  & \textbf{3.33}  & 9.65  & \textbf{8.73} \\
        \midrule
        V1-R        & ReLU      & NPE    & 3.95  & 3.54  & 9.85  & 9.07 \\ 
        V1-T        & Tanh      & NPE    & 3.87  & 3.47  & 9.77  & 8.87 \\
        V1-S        & Sigmoid   & NPE    & 3.89  & 3.44  & 9.59  & 8.75 \\
        V1-E        & ELU       & NPE    & \textbf{3.75}  & \textbf{3.36}  & \textbf{9.35}  & \textbf{8.64} \\
        \bottomrule
    \end{tabular}
    \begin{tablenotes}\footnotesize
        \item \textbf{[L]MLA}: [\textbf{L}earnable] \textbf{M}ultiplicative \textbf{L}inear \textbf{A}ttention; \\
    \textbf{V\{0/1\}}: Linear Attention \textbf{V}ariations with \{Additive RPE / No PE\}\\
    \textbf{*-\{R/T/S/E\}}: LA block with $\psi_{\{relu/tanh/sigmoid/elu\}}(\cdot)$ \\
    \end{tablenotes}
    \vspace{-4mm}
\end{table}

activation function $\psi_{elu}(\cdot)$ and LM-APE in LMLA block.

The results are illustrated in Table\ref{tb:la-compare}.
Here, all the number of head in encoder is set to 8 and GLU is not used for fair.
Next, we will take the decoding results of attention rescoring as an example to analyze the key points we put forward.

Under four different linear attention paradigm, $\psi_{elu}(\cdot)$ obtained an average of 3.35/8.59\% on test-clean/test-other.
About the other three, $\psi_{relu}(\cdot)$ obtained average 3.43/9.00\%, $\psi_{tanh}(\cdot)$ achieves 3.42/8.85\%, $\psi_{sigmoid}(\cdot)$ achieves 3.36/8.73\%.
Obviously, $\psi_{elu}(\cdot)$ is the best choice in our design.

In comparison, our proposed LMLA performs better than other variations with different styles of positional embedding, which achieves 3.30/8.74\% on average.
Otherwise, MLA performs an average of 3.42/8.74\%, which shows the effectiveness of learnable embeddings.
V0 with A-RPE gets the results of 3.38/8.86\%.
Compared with APE, we can see that the advantage of RPE in linear attention is not apparent as usual.
Furthermore, V1 gets 3.45/8.83\%, which further reveals that the absence of positional embedding will bring specific degradation effects.

On the other hand, we can get the same conclusions in terms of the ctc greedy search scenario.

\vspace{-3mm}
\subsubsection{Effect of Gated Linear Units}


\begin{table}[htbp]\small
    \vspace{-4mm}
    \centering
    \caption{\label{tb:glu} Effect of GLU}
    \begin{tabular}{cC{3.5mm}C{8mm}C{5mm}C{8.5mm}C{5mm}C{8.5mm}}
        \hline
        \toprule \\
        \specialrule{0em}{-10pt}{1pt}
        \multirow{2}{*}[-3mm]{Model} & \multirow{2}{3.5mm}[-3mm]{with GLU} & \multirow{2}{8mm}[-1mm]{GLU Acti-vation} & \multicolumn{2}{c}{Test Clean} & \multicolumn{2}{c}{Test Other} \\
        \cmidrule{4-7}  &  &  &  ctc greedy & attention rescore & ctc greedy & attention rescore   \\
        \midrule
        Conformer   & $\times$      & -       & \textcolor{blue}{3.51}  & \textcolor{blue}{3.31}  & \textcolor{blue}{9.12}  & \textcolor{blue}{8.51} \\
        \midrule
        LMEC-E      & $\times$      & -       & 3.70  & 3.32  & 9.37  & 8.47 \\
        \midrule
        LMEC-E-$S_w$    & $\checkmark$  & Swish   & 3.61  & 3.24  & 9.34  & 8.50 \\
        LMEC-E-E        & $\checkmark$  & ELU     & 3.70  & 3.24  & 9.24  & 8.41 \\
        LMEC-E-R        & $\checkmark$  & ReLU    & 3.62  & 3.30  & \textbf{9.14}  & 8.39 \\
        LMEC-E-G        & $\checkmark$  & GeLU    & \textbf{3.57}  & \textbf{3.17}  & 9.18  & \textbf{8.38} \\
        \bottomrule
    \end{tabular}
\end{table}

Then, we test the effectiveness of the GLU module and the result is illustrated in Table\ref{tb:glu}.
In this set of comparative experiments, the head number of all models is set to 8 as well.

In Table\ref{tb:glu}, four kinds of GLU modules with different activation functions are evaluated.
It can be seen that when applying the GLU module, the WER of our proposed LMEC model can be reduced, and LMEC-E-G with GeLU activation function achieves the best results.
Specifically, when we use ctc greedy search to evaluate the models, LMEC-E-G achievements 3.51/2.03\% relative WER reduction on test-clean/test-other compared with LMEC-E without GLU module.
Meanwhile, through the attention rescoring, LMEC-E-G outperforms LMEC-E 4.52/1.06\% relative WER reduction on test-clean/test-other.
In addition, our final proposed LMLA-E-G obtains a comparable WER compared with Conformer.

\vspace{-2mm}
\subsection{Effect of Training and Inference Method}

\begin{table}[htbp]\small
    \vspace{-6mm}
    \centering
    \caption{\label{tb:lr-prod} Differentiated Training Method}
    \begin{tabular}{ccC{5mm}C{8.5mm}C{5mm}C{8.5mm}}
        \hline
        \toprule \\
        \specialrule{0em}{-10pt}{1pt}
        \multirow{2}{*}[-3mm]{Model} & \multirow{2}{*}[-3mm]{Train Method} & \multicolumn{2}{c}{Test Clean} & \multicolumn{2}{c}{Test Other} \\
        \cmidrule{3-6}  &  &  ctc greedy & attention rescore & ctc greedy & attention rescore   \\
        \midrule \\
        \specialrule{0em}{-10pt}{1pt}
        LMEC-E      & L-Prod      & \textbf{3.70}  & \textbf{3.32}  & \textbf{9.37}  & \textbf{8.47} \\
        LMEC-E      & R-Prod      & 3.81  & 3.34  & 9.54  & 8.69 \\
        \midrule \\
        \specialrule{0em}{-10pt}{1pt}
        LMEC-E-G    & L-Prod      & \textbf{3.57}  & \textbf{3.17}  & \textbf{9.18}  & \textbf{8.38} \\
        LMEC-E-G    & R-Prod      & 3.61  & 3.17  & 9.26  & 8.45 \\
        \bottomrule
    \end{tabular}
     \begin{tablenotes}\footnotesize
        \item \textbf{[L]MLA}-*-\{$S_w$/E/R/G\}: \textbf{GLU} block with different activation functions.
      \end{tablenotes}
\end{table}

It is not difficult to see from the Table\ref{tb:lr-prod} that compared with R-Prod training, L-Prod training effect and model convergence stability are better.
Because the calculation result of L-Prod and R-Prod are equal in the forward pass of cosFormer, LBLA and LMLA, the comparison results in this paper are all obtained by L-Prod training and R-Prod inference.


\vspace{-2mm}
\section{Conclusions}
This paper demonstrates that multiplicative RPE is not cost-effective in ASR tasks under the paradigm of linear attention, and it is recommended to use APE that can be learned to replace cosFormer kernel.
Furthermore, the authors recommend using GLU module with GeLU activation function to replace FFN to further improve model performance.
In order to illustrate the effectiveness of the proposed points, the authors mainly conduct experiments on the encoder part, and in fact, the innovative points of the experiments is completely orthogonal to the decoder part.

\bibliographystyle{IEEEbib}
\bibliography{cos-conformer-yyg}
\end{document}